\documentclass[aps,twocolumn,showpacs,preprintnumbers,amsmath,amssymb,superscriptaddress,floatfix,nofootinbib]{revtex4}

\usepackage{graphicx}
\usepackage{hyperref}
\usepackage{amsmath}
\allowdisplaybreaks[1]
\usepackage{amsfonts}
\usepackage{amssymb}
\usepackage{multirow}
\usepackage{makecell}
\usepackage[dvipsnames]{xcolor}
\usepackage{booktabs}
\usepackage{tabularx}

\begin{document}

\title{Prediction of five-flavored pentaquarks}

\author{Chao-Wei Shen} \email{c.shen@fz-juelich.de}
\affiliation{Institute for Advanced Simulation, Institut f\"ur Kernphysik and
J\"ulich Center for Hadron Physics, \\ Forschungszentrum J\"ulich, D-52425 J\"ulich, Germany}

\author{Ulf-G. Mei\ss{}ner} \email{meissner@hiskp.uni-bonn.de}
\affiliation{Helmholtz-Institut f\"ur Strahlen- und Kernphysik and Bethe Center for
Theoretical Physics,\\ Universit\"at Bonn, D-53115 Bonn, Germany}
\affiliation{Institute for Advanced Simulation, Institut f\"ur Kernphysik and J\"ulich
Center for Hadron Physics, \\ Forschungszentrum J\"ulich, D-52425 J\"ulich, Germany}
\affiliation{Tbilisi State University, 0186 Tbilisi, Georgia}

\date{\today}

\begin{abstract}
We analyze the possibility of the existence of a fully exotic pentaquark state
$udsc\bar{b}$, which is made from the five different flavors participarting in the
strong interactions.
We investigate the coupled channel effects of the $B^{(*)} \Xi_c-B^{(*)} \Xi_c^\prime$ system
through  $t$-channel vector meson exchange to search for such states.
A $B \Xi_c$ and a $B \Xi_c^\prime$ bound state are found to have binding energies
of about $10-30$~MeV with $B_s\Lambda_c$ and $B_c\Lambda$ being the possible decay channels.
Similarly, a bound state is formed in the $B^* \Xi_c$  and $B^* \Xi_c^\prime$ channels, respectively.
These states could be searched for through the $pp\to B_c\Lambda X$ or
$pp\to B_s^{(*)}\Lambda_c X$ processes by the LHCb collaboration.

\end{abstract}

\maketitle

{\em Introduction:}
In Refs.~\cite{Wu:2010jy,Wu:2010vk}, Wu et al. predicted
several $N_{c\bar{c}}$ and $\Lambda_{c\bar{c}}$ states in a vector-meson exchange model.
The $N_{c\bar{c}}$ states are considered as  S-wave $\bar{D}^{(*)} \Sigma_c$ bound states (hadronic
molecules), and such kind of particles were observed in the $\Lambda_b^0 \to J/\psi K^- p$ process
by the LHCb Collaboration in 2015~\cite{LHCb:2015yax} and further scrutinized in 2019 with  Run-II
data~\cite{LHCb:2019kea}.
In 2020, a peak structure was observed in the $\Xi_b^- \to J/\psi K^- \Lambda$ decay by the
LHCb Collaboration with a significance of $3.1\sigma$~\cite{LHCb:2020jpq}.
It is regarded as a hidden-charm pentaquark with strangeness, which is consistent with the
predicted $\Lambda_{c\bar{c}}$ state, that is, a $\bar{D}^{(*)} \Xi_c^{(\prime)}$ bound state~\cite{Chen:2020uif,Peng:2020hql,Chen:2020kco,Liu:2020hcv,Xiao:2021rgp}.
These hidden-charm pentaquark states soon attracted much attention and became a very hot
topic in particle physics.
Similar methods were applied to the hidden beauty case~\cite{Wu:2010rv,Shen:2017ayv} as well.
These states still await experimental scrutiny.
Reviews of such  pentaquarks (and other exotic hadrons) are given in
Refs.~\cite{Chen:2016qju,Richard:2016eis,Lebed:2016hpi,Guo:2017jvc,Ali:2017jda,Liu:2019zoy,Yang:2020atz}.
%
The doubly charmed pentaquark states also attracted much attention, especially
with the observation of a doubly charmed tetraquark $T_{cc}^+$ by the LHCb
Collaboration~\cite{LHCb:2021vvq,LHCb:2021auc}, see e.g. Refs.~\cite{Guo:2011dd,Chen:2017vai,Guo:2017vcf,Shimizu:2017xrg,Zhou:2018bkn,Dias:2018qhp,Yan:2018zdt,Wang:2018lhz,Park:2018oib,Zhu:2019iwm,Yu:2019yfr,Yang:2020twg,Chen:2021htr,Dong:2021bvy,Chen:2021kad}.
Another development of relevance to our work was the prediction of an  S-wave $B\bar{D}$ bound state
with isospin $I=0$ in the effective Lagrangian framework~\cite{Zhang:2006ix}.
Inspired by this work together with the explorations for the hidden-charm and double-charm
hadronic molecules, the charm-antibeauty pentaquark states, which also belongs to the
heavy-antiheavy system, are also expected to exist, and some are considered in Ref.~\cite{Peng:2019wys}.
Here, we will provide theoretical evidence of the existence of
the $B^{(*)} \Xi_c^{(\prime)}$ bound states with charge $Q=+1$.
On the one hand, the $B^{(*)} \Xi_c-B^{(*)} \Xi_c^\prime$ system is  most likely to form bound
states among several possible options.
On the other hand, it is quite special that all the involved quarks have different flavors,
since the quark content is $[q\bar{b}][q^\prime sc]$($q^{(\prime)}=u, d$).
Thus, this system is  explored in the present work and it appears to be  the best option
to search for a pentaquark state, which we call $\Lambda_{c\bar{b}}$, that includes five
different quarks.
Moreover, the properties including their pole positions, decay widths and couplings to other
channels are extracted.
This can definitely make the hadronic molecule spectrum more complete~\cite{Dong:2021bvy}.
%

\begin{figure}[b!]
	\centering
	\includegraphics[width=0.3\linewidth]{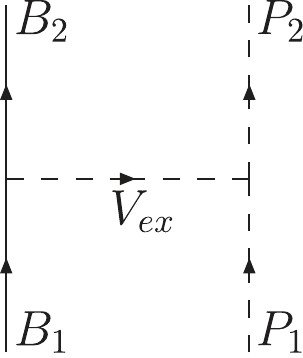} \ \ \ \ \ \ \ \ \
	\includegraphics[width=0.3\linewidth]{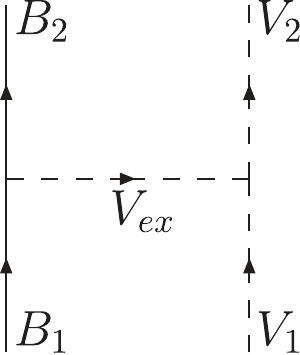}
	\caption{Feynman diagrams for the pseudoscalar meson-baryon interaction (left)
          and vector meson-baryon interaction (right) via the exchange of a vector meson in
          the $t$-channel. $P_1$ and $P_2$ are $B$, $B_s$ and $B_c$, $V_1$ and $V_2$ are $B^*$ and
          $B_s^*$, $B_1$ and $B_2$ are $\Xi_c$, $\Xi_c^\prime$, $\Lambda_c$ and $\Lambda$ and
          $V_{ex}$ is $\rho$, $\omega$, $\bar{K}^*$, $D^*$ and $D_s^*$.}
	\label{Fig:FD1}
\end{figure}

{\em Theoretical Framework:}~In this work, we investigate the coupled channel effects in
the $B^{(*)} \Xi_c-B^{(*)} \Xi_c^\prime$ system following the method in Ref.~\cite{Wu:2010jy} to
search for possible bound states with $Q=+1$.
The interaction is given via the exchange of vector mesons in the $t$-channel. For more details,
see~\cite{SM}.
The calculation is divided into two parts, namely the pseudoscalar meson-baryon ($PB$) interaction and
the vector meson-baryon ($VB$) interaction.
For the $B \Xi_c-B \Xi_c^\prime$ sector, there are in total four channels taken into consideration,
$B\Xi_c$, $B\Xi_c^\prime$, $B_s\Lambda_c$ and $B_c\Lambda$.
For the $B^* \Xi_c-B^* \Xi_c^\prime$ sector, since no $B_c^*(J^P=1^-)$ state has been found so far,
only three channels $B^*\Xi_c$, $B^*\Xi_c^\prime$ and $B_s^*\Lambda_c$  are considered,
and we will focus on these three channels in the $VB$ case.
The corresponding Feynman diagrams of these two systems are presented in Fig.~\ref{Fig:FD1}.

Then the unitary scattering amplitude $T$ can be derived from the coupled-channel Bethe-Salpeter
equation in the on-shell factorization approach of Refs.~\cite{Oset:1997it,Oller:2000fj}:
\begin{equation}
  T = [1-VG]^{-1}V~.
\end{equation}
Details on the interaction kernel $V$, the meson-baryon loop function $G$ and the pertinent
regularization are given in~\cite{SM}.
%
%
Possible bound states are given by poles in the complex plane on different Riemann sheets,
which we are searching. For a bound state or a resonance, its coupling strength to different channels
can be derived from the residues $a_{-1,ab}$, which appear in the Laurent expansion of
$T^{II}_{ab}$~\cite{Ronchen:2012eg}:
\begin{equation}
  T^{II}_{ab}(z) = \frac{a_{-1,ab}}{z-z_0} + {\cal R}~, 
\end{equation}
where $a_{-1,ab} = g_a g_b$, ${\cal R}$ is the nonsingular remainder and the superscript $II$
refers to the second Riemann sheet.
However, a global phase is always undetermined.
Once we find a pole $z_0$ in $T$, we label the closest channel as $a$ and this global sign is determined from $g_a=\sqrt{a_{-1,aa}}$.
Then the couplings with other channels $g_b$ are obtained from the residues $a_{-1,ab}$.

\begin{figure}[t!]
\includegraphics[width=0.87\linewidth]{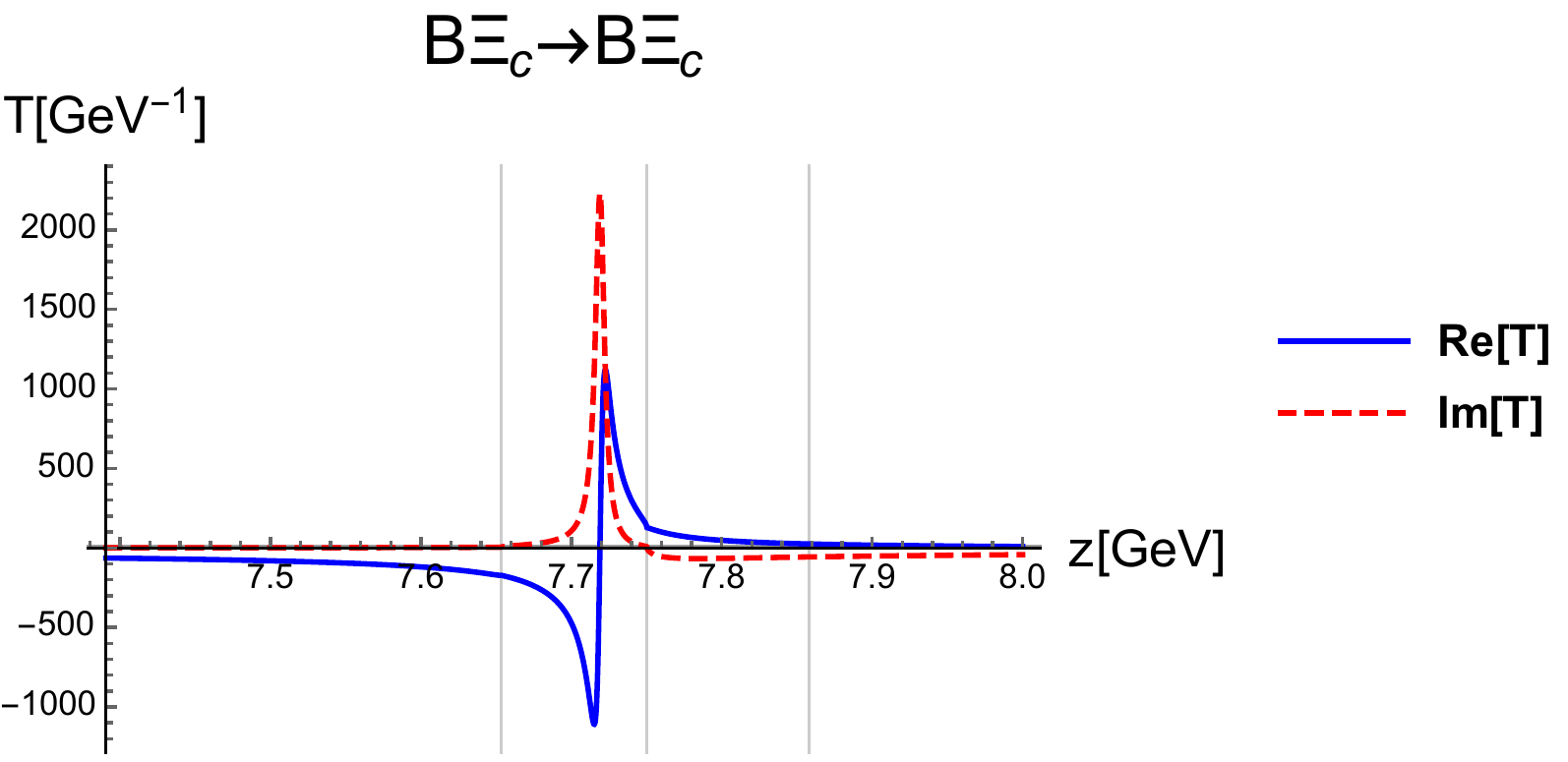}
\includegraphics[width=0.87\linewidth]{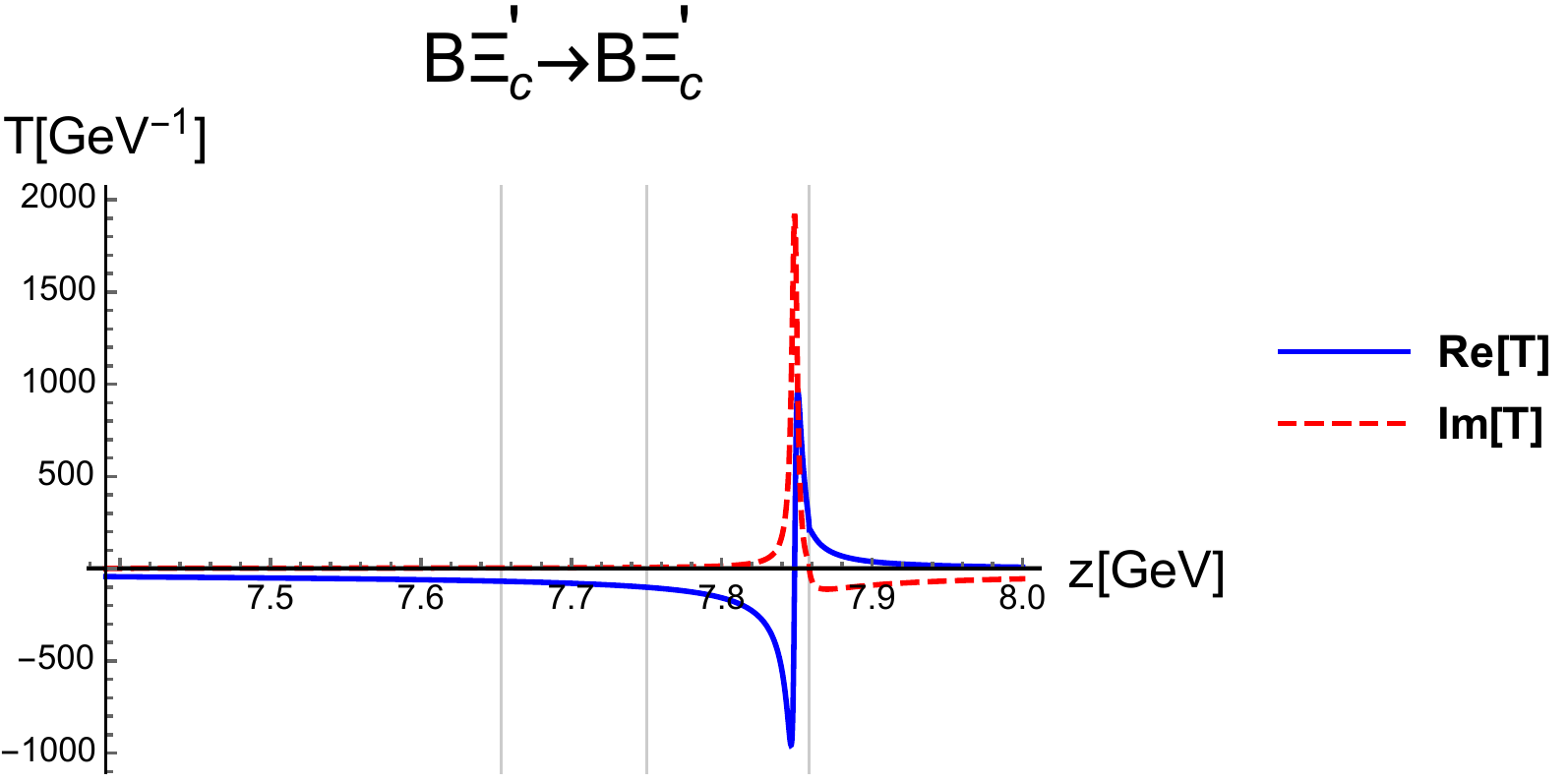}
\caption{The real (blue solid lines) and imaginary (red dashed lines) part of the
amplitudes $T$ of $B\Xi_c \to B\Xi_c$ (upper panel) and $B\Xi_c^\prime \to B\Xi_c^\prime$
(lower panel) with
$\mu=1$~GeV and $a(\mu)=-3.1$. The three thin solid lines are the thresholds of $B_s\Lambda_c$,
$B\Xi_c$ and $B\Xi_c^\prime$, in order.  \label{Fig:RI1}}
\end{figure}

\begin{figure}[htbp]
	\includegraphics[width=0.87\linewidth]{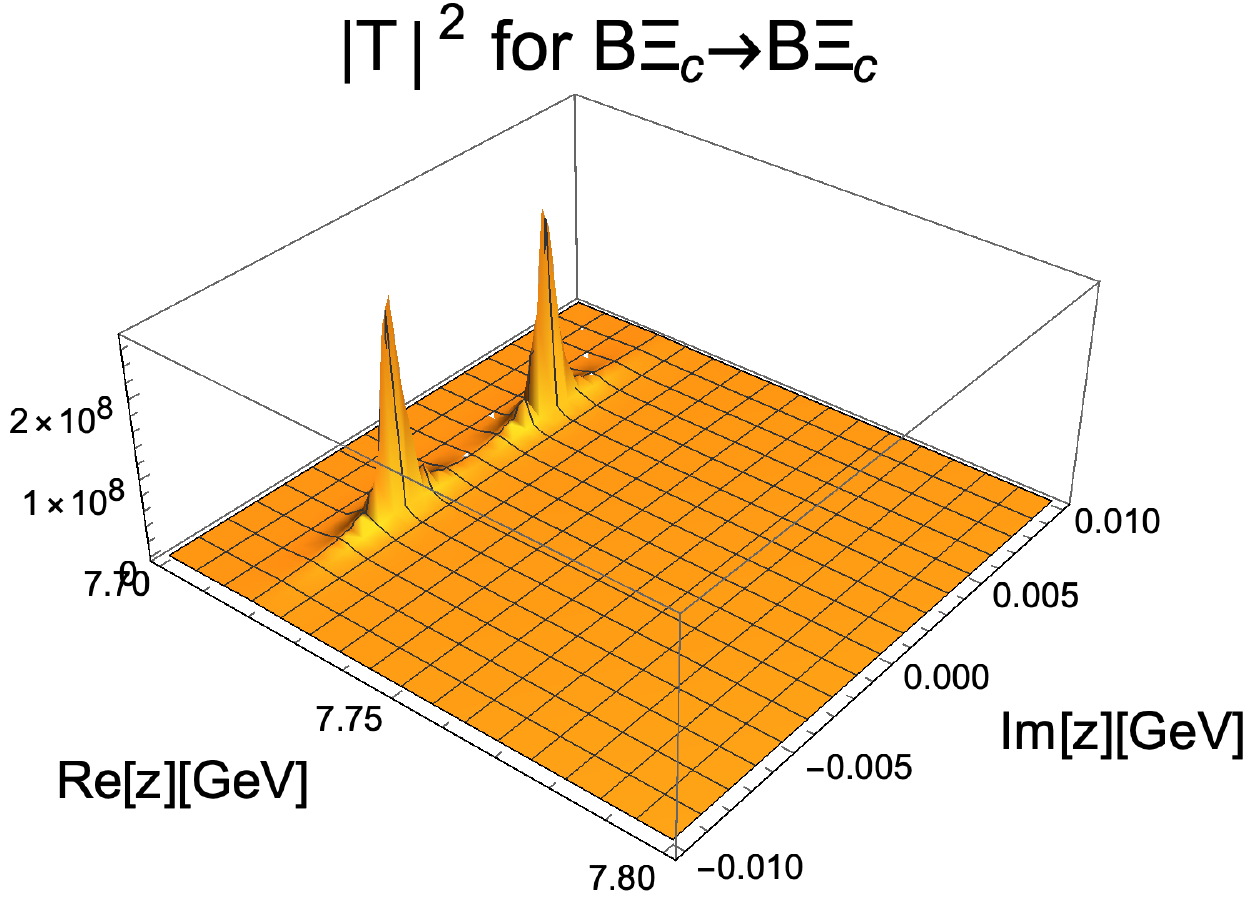}
	\includegraphics[width=0.87\linewidth]{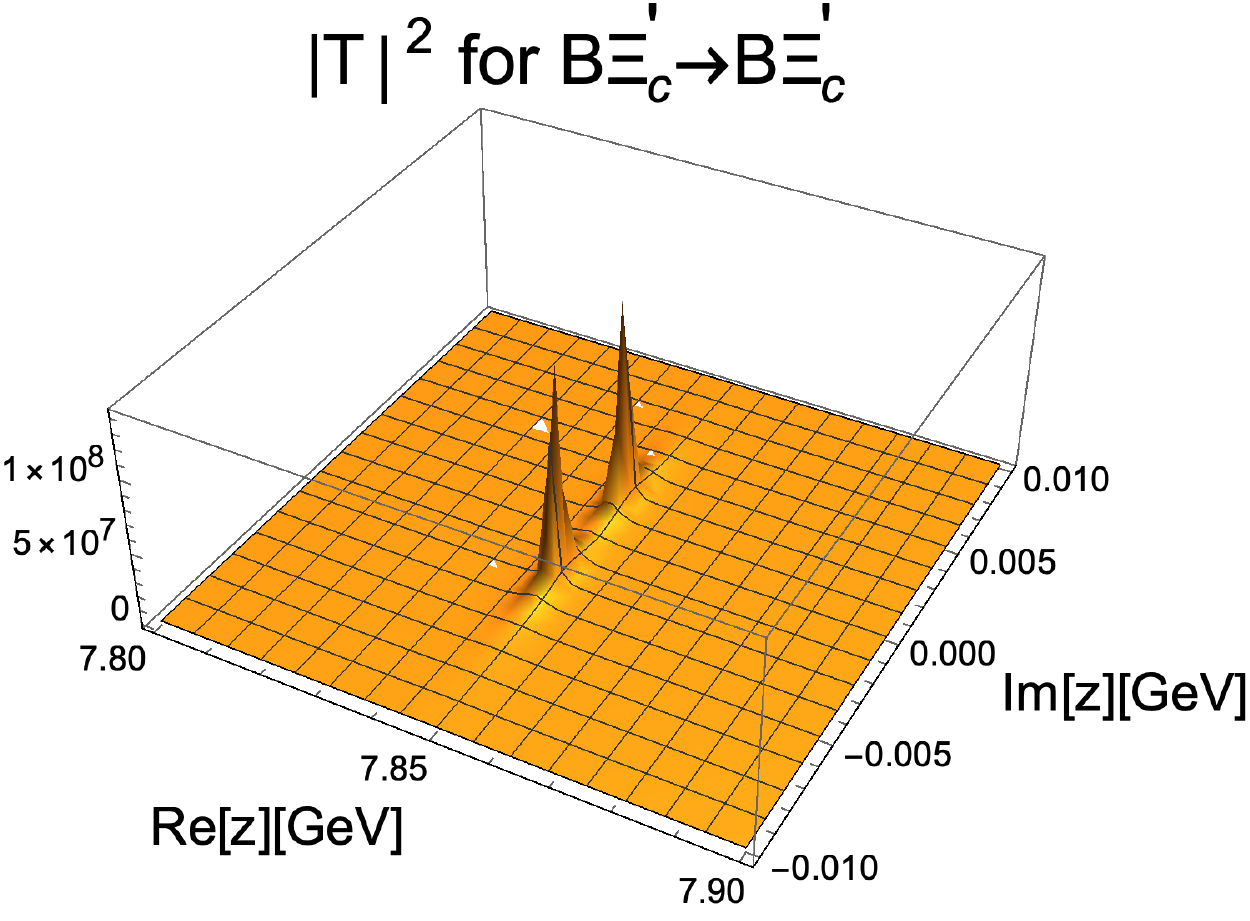}
	\caption{The amplitude squared $|T|^2 [\rm{GeV}^{-2}]$ of $B\Xi_c \to B\Xi_c$ (upper panel)
          and $B\Xi_c^\prime \to B\Xi_c^\prime$ (lower panel) in the complex energy $z$ plane
          with $\mu=1$~GeV and $a(\mu)=-3.1$. \label{Fig:3D1}}
\end{figure}

{\em Results and discussion:}~~We first investigate the pseudoscalar meson-baryon interaction
with two possible isospins $I=0$ and $I=1$, that is the $B \Xi_c-B \Xi_c^\prime$ sector.
Since we seek bound states in the $B\Xi_c$ and $B\Xi_c^\prime$ channels, whose thresholds
are at $7750.19$~MeV and $7858.12$~MeV, respectively. First, we consider these two channels
separately to make an exploratory calculation.
The pole positions $z_0$ and coupling constants $g_a$ of the states generated from $B\Xi_c$ and
$B\Xi_c^\prime$ channels with $a(\mu=1~\rm{GeV})=-3.1$ for $I=0$ and $1$ are given in
Tab.~\ref{Table:re1}, see~\cite{SM} for details.
\begin{table}[htbp]
\centering
\renewcommand\arraystretch{1.2}
\caption{Pole positions $z_0$ and coupling constants $g_a$ for the states generated from
the $B\Xi_c$ and $B\Xi_c^\prime$ channels with $\mu=1$~GeV and $a(\mu)=-3.1$. \label{Table:re1}}
\begin{tabular}{p{1.7cm}<{\centering}p{1.9cm}<{\centering}p{1.7cm}<{\centering}p{1.7cm}<{\centering}}
\hline
\hline
\multirow{2}{*}{Isospin} & \multirow{2}{*}{$z_0$ [MeV]}	&	\multicolumn{2}{c}{$g_a$} \\
&	&	$B\Xi_c$ & $B\Xi_c^\prime$	\\
\hline
0  & 7740.07  & 2.141	&	0 \\
	&	7849.39	&	0	&	2.015	\\
\hline
\hline
\end{tabular}
\end{table}
We find that for $I=0$ there exists a $B\Xi_c$ bound state and a $B\Xi_c^\prime$ bound state.
It can be seen in Tab.~\ref{Table:IF0} in~\cite{SM} that $B\Xi_c$ and $B\Xi_c^\prime$ are independent
when we only consider these two channels.
This is actually a calculation of two single channels, and as a result the bound state couples
only with each corresponding channel.
Clearly,  after adding more channels, such as $B_s\Lambda_c$ and $B_c\Lambda$, these two channels
are coupled and the generated states can interact with all the related channels.
For $I=1$, there is no bound state in both cases.
This can be understood from the fact that the signs of $C_{ab}$ for $I=0$ and $1$ are different, the
transition potential for $I=1$ is repulsive, see~\cite{SM}.
Thus, we do not discuss  the isospin $1$ case any longer,
and only focus on the isospin $I=0$ case in what follows.
%
%
Now we take all the four possible channels $B\Xi_c$, $B\Xi_c^\prime$, $B_s\Lambda_c$ and $B_c\Lambda$
into account.
The peak positions of $T$ on the real axis are found around $7719.04$~MeV and $7848.43$~MeV.
The real and imaginary part of the amplitudes $T$ for these two poles in $B\Xi_c \to B\Xi_c$
and $B\Xi_c^\prime \to B\Xi_c^\prime$ are presented in Fig.~\ref{Fig:RI1}.
It is found that the two peak structures are located just below the threshold of the
$B\Xi_c$ and $B\Xi_c^\prime$ channels, making them candidates for hadronic molecules.
Next, we turn to the case of complex energy $z$.
The amplitudes squared $|T|^2$ of $B\Xi_c \to B\Xi_c$ and $B\Xi_c^\prime \to B\Xi_c^\prime$ in the
complex energy plane are shown in Fig.~\ref{Fig:3D1}.
The pole positions $z_0$ and coupling constants $g_a$ for the dynamically generated
$I=0$ states with $a(\mu=1~\rm{GeV})=-3.1$ are given in Tab.~\ref{Table:re2}.

\begin{table*}[htbp]
\centering
\renewcommand\arraystretch{1.2}
\caption{Pole positions $z_0$ and coupling constants $g_a$ for the $I=0$ states generated
in the $PB$ sector with $\mu=1$~GeV and $a(\mu)=-3.1$. \label{Table:re2}}
\begin{tabular}{p{2.6cm}<{\centering}p{2.6cm}<{\centering}p{2.6cm}<{\centering}p{2.6cm}<{\centering}p{2.6cm}<{\centering}}
\hline
\hline
\multirow{2}{*}{$z_0$ [MeV]}	&	\multicolumn{4}{c}{$g_a$} \\
&	$B\Xi_c$ & $B\Xi_c^\prime$ & $B_s\Lambda_c$ & $B_c\Lambda$	\\
\hline
$7720.12-3.7i$	&	$2.851+0.100i$	& $-0.0003-0.076i$ & $0.294-0.371i$ & $-0.333+0.059i$ \\
&	2.853	&	0.076	&	0.473	&	0.338	\\
$7847.60-2.2i$	& $0.028+0.015i$& $2.126+0.124i$&	$0.005-0.002i$	& $0.453-0.064i$	\\
&	0.032	&	2.129	&	0.005	&	0.458	\\
\hline
\hline
\end{tabular}
\end{table*}

We find  two poles in the complex energy plane and they are responsible for
the peak structures observed on the real axis.
These two poles, which are just below the thresholds of $B\Xi_c$ and $B\Xi_c^\prime$, are in
accordance with the result of the two channels.
The one at $z_0=7720.12-3.7i$~MeV couples much stronger to $B\Xi_c$ than to other channels.
This indicates that it is a $B\Xi_c$ bound state with  binding energy of $30.1$~MeV.
The other pole is located at $z_0=7847.60-2.2i$~MeV, it has the strongest coupling with the
$B\Xi_c^\prime$ channel, thus we  regard it as a $B\Xi_c^\prime$ bound state. Its
binding energy is $10.5$~MeV.
Because such $B \Xi_c^{(\prime)}$ bound states can not decay to light meson-light baryon channels
via the strong interaction, the most relevant two-body decay channels are just those we considered.
The pertinent widths of these two states to $B_s\Lambda_c$ and $B_c\Lambda$  are about $7.4$~MeV and
$4.4$~MeV, respectively.
Note that these two states will be more loosely bound when $a(\mu)$ takes a smaller value, that is, the
peak structures will gradually move away from the threshold. Explicit results for different values of
the subtraction constant $a(\mu)$ are given in~\cite{SM}.

We now consider the vector meson-baryon interaction in this section, that is the
$B^* \Xi_c-B^* \Xi_c^\prime$ sector.
We found that in the $PB$ case  the potential with $I=1$ is repulsive and
no bound states can be formed.
This feature is also present in the $VB$ sector, so we only consider the
isospin $0$ case here.
As mentioned before, there are only three channels in this case.
Again, we first calculate with the two channels $B^*\Xi_c$ and $B^*\Xi_c^\prime$ only,
whose thresholds are at $7795.52$~MeV and $7903.45$~MeV, respectively, to check for
the existence of possible bound states.
The pole positions $z_0$ and coupling constants $g_a$ for the $I=0$ states generated from the
$B^*\Xi_c$ and $B^*\Xi_c^\prime$ channels with $a(\mu=1~\rm{GeV})=-3.1$ are
listed in Tab.~\ref{Table:re3}.
The result turns out to be similar to the one in the $PB$ case, that is, a bound state is
found in each channel.

\begin{figure}[htbp]
\includegraphics[width=0.87\linewidth]{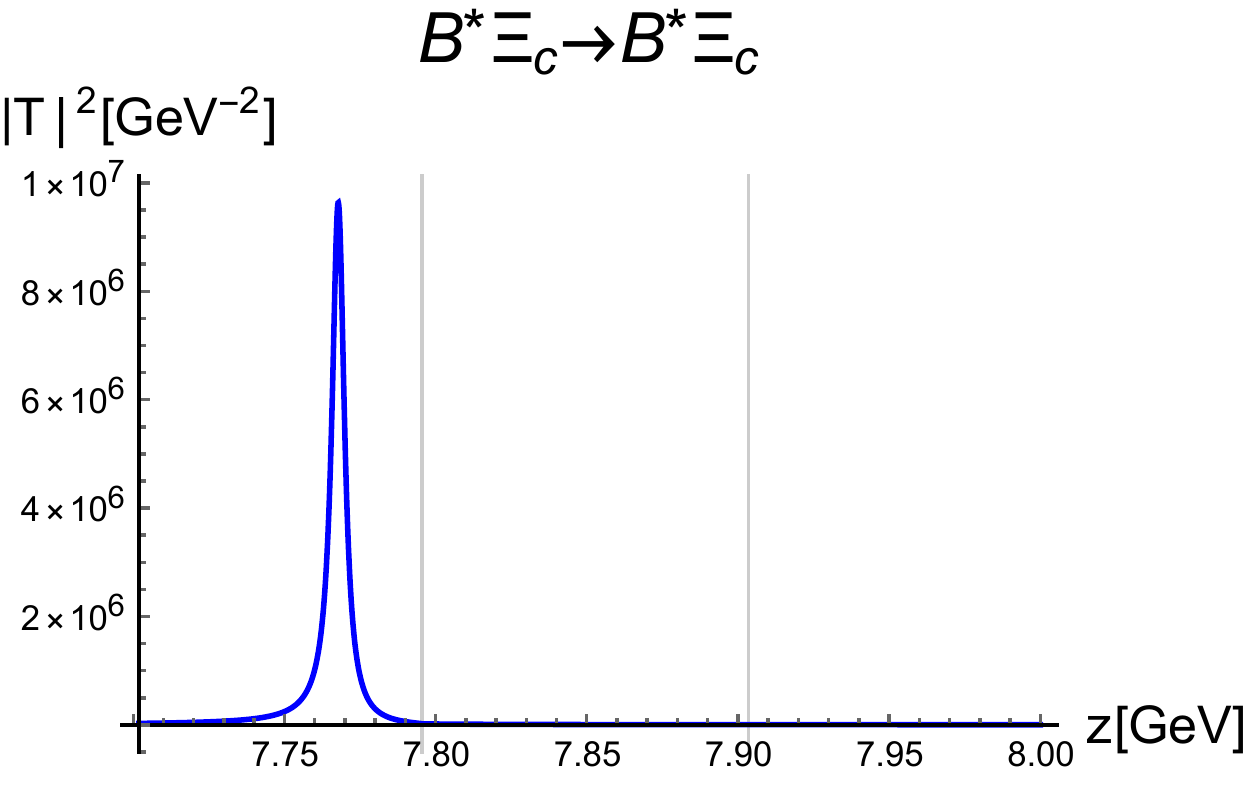}
\includegraphics[width=0.87\linewidth]{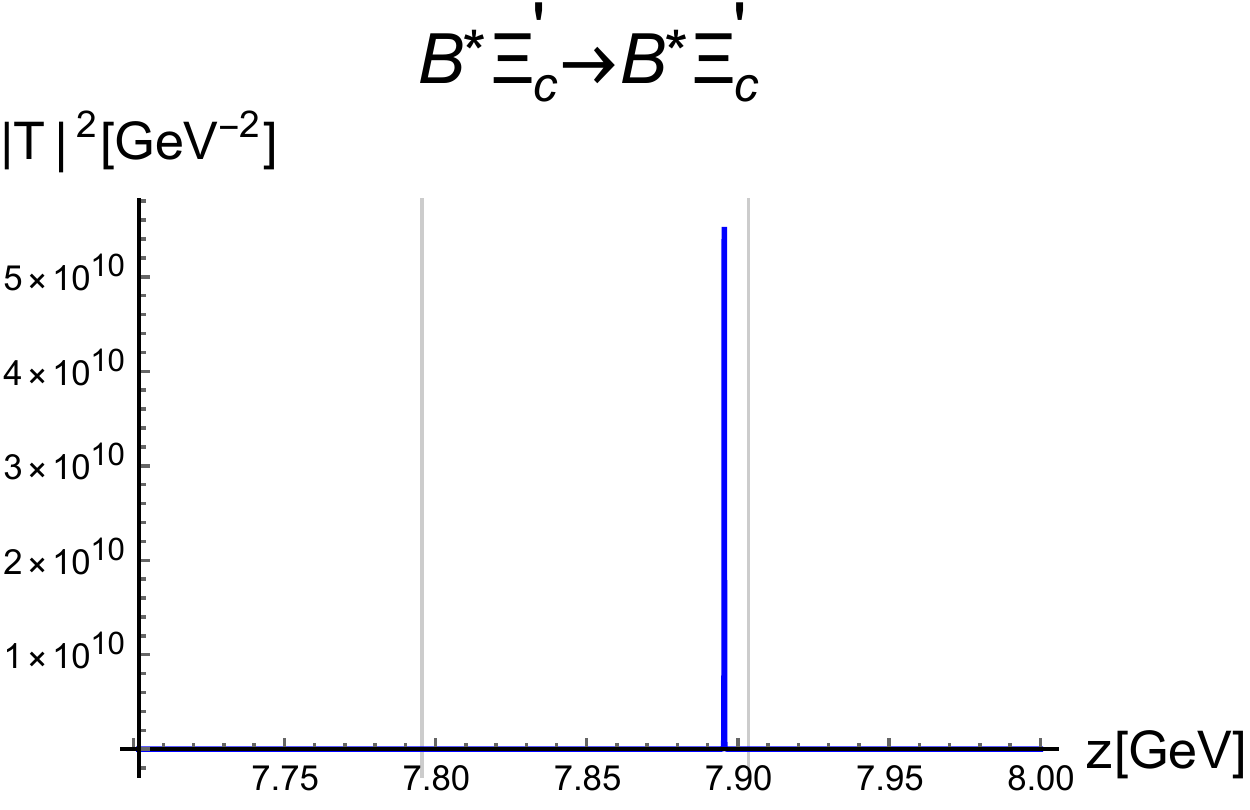}
\caption{The amplitude squared $|T|^2$ of $B^*\Xi_c \to B^*\Xi_c$ (upper panel)
and $B^*\Xi_c^\prime \to B^*\Xi_c^\prime$ (lower panel) with $\mu=1$~GeV and $a(\mu)=-3.1$.
The two thin solid lines are the thresholds of $B^*\Xi_c$ and $B^*\Xi_c^\prime$, in order.
\label{Fig:T2}}
\end{figure}

\begin{figure}[htbp]
\includegraphics[width=0.87\linewidth]{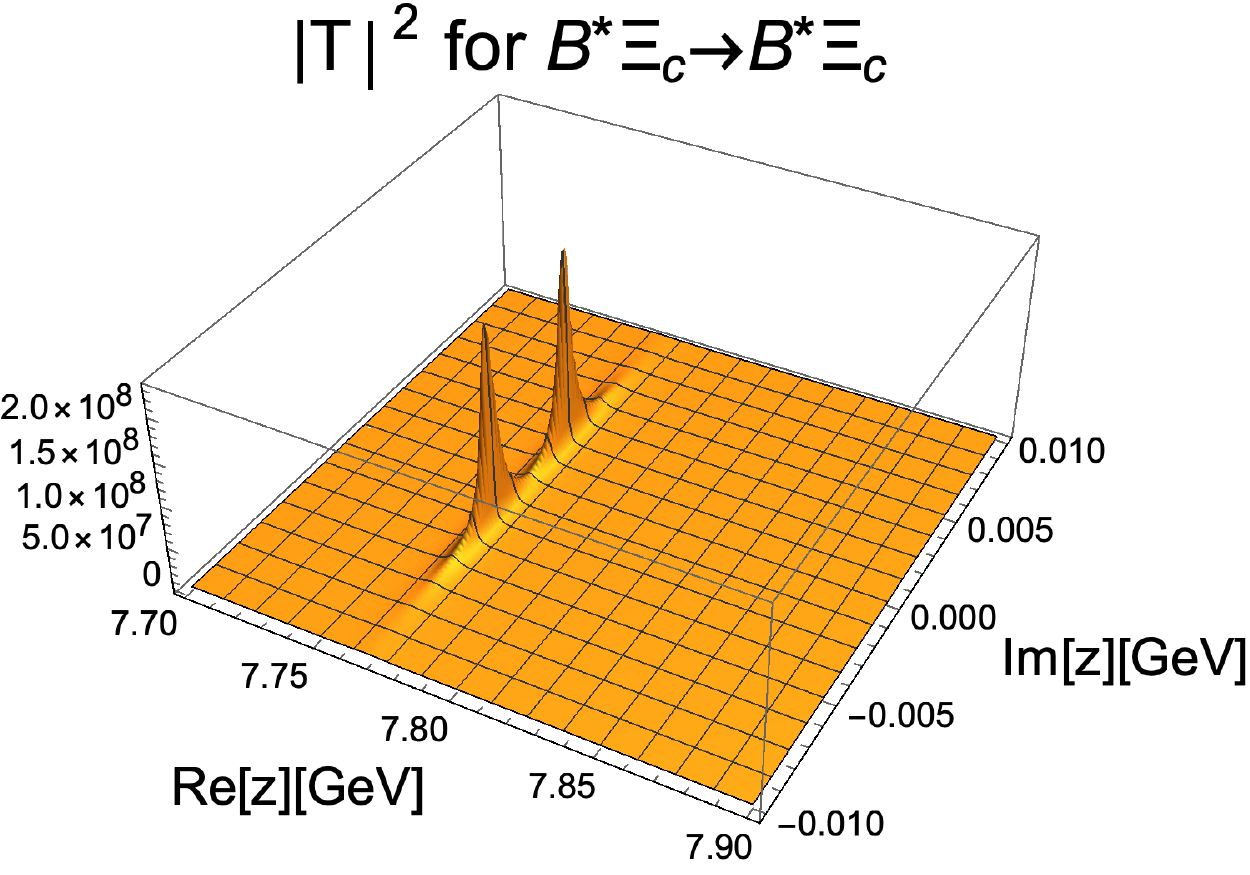}
\includegraphics[width=0.87\linewidth]{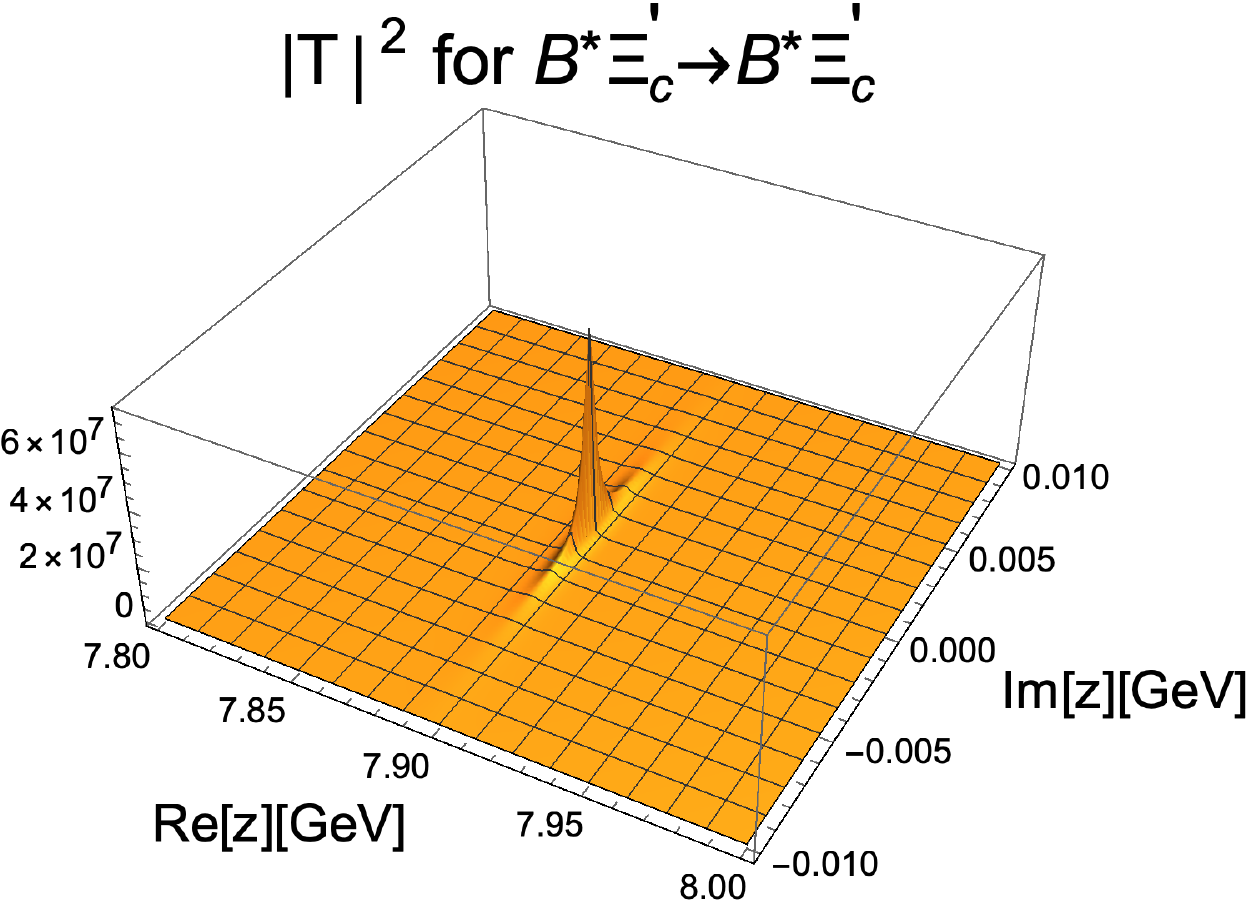}
\caption{The amplitude squared $|T|^2 [\rm{GeV}^{-2}]$ of $B^*\Xi_c \to B^*\Xi_c$ (upper panel) and
$B^*\Xi_c^\prime \to B^*\Xi_c^\prime$ (lower panel) in the complex energy $z$ plane with
$\mu=1$~GeV and $a(\mu)=-3.1$.
\label{Fig:3D2}}
\end{figure}

\begin{table}[htbp]
\centering
\renewcommand\arraystretch{1.2}
\caption{Pole positions $z_0$ and coupling constants $g_a$ for the $I=0$ states generated
from the $B^*\Xi_c$ and $B^*\Xi_c^\prime$ channels with $\mu=1$~GeV and $a(\mu)=-3.1$.
\label{Table:re3}}
\begin{tabular}{p{1.9cm}<{\centering}p{1.7cm}<{\centering}p{1.7cm}<{\centering}}
\hline
\hline
\multirow{2}{*}{$z_0$ [MeV]}	&	\multicolumn{2}{c}{$g_a$} \\
&	$B^*\Xi_c$ & $B^*\Xi_c^\prime$	\\
\hline
7785.25 & 2.154	&	0 \\
7894.43	&	0	&	2.037	\\
\hline
\hline
\end{tabular}
\end{table}
After confirming the existence of the two bound states, we calculate all the
three channels $B^*\Xi_c$, $B^*\Xi_c^\prime$ and $B_s^*\Lambda_c$.
As can be read off from Tab.~\ref{Table:IF02} in~\cite{SM}, the $B^*\Xi_c^\prime$ channel
is still independent from the other two, while $B^*\Xi_c$ and $B_s^*\Lambda_c$ are coupled.
Thus, to the order we are working, we actually deal with  a two channel and a single channel
calculation.
The amplitudes squared $|T|^2$ of $B^*\Xi_c \to B^*\Xi_c$ and $B^*\Xi_c^\prime \to B^*\Xi_c^\prime$
are shown in Fig.~\ref{Fig:T2}.
A peak with a certain width around $7767.80$~MeV can be seen in the $B^*\Xi_c \to B^*\Xi_c$ reaction.
In $B^*\Xi_c^\prime \to B^*\Xi_c^\prime$, the structure located at 7894.43~MeV corresponds to a pole
and displays no width.
This is because $B^*\Xi_c^\prime$ is decoupled and has no interactions with any involved channels
in the calculation.
However, once any possible decay channels are included into the calculation, it would
behave just like that in the $B^*\Xi_c \to B^*\Xi_c$ process.
The amplitudes squared $|T|^2$ of $B^*\Xi_c \to B^*\Xi_c$ and $B^*\Xi_c^\prime \to B^*\Xi_c^\prime$
in the complex energy plane are shown in Fig.~\ref{Fig:3D2}.
The pole positions $z_0$ and coupling constants $g_a$ for the dynamically generated $I=0$
states with $a(\mu=1~\rm{GeV})=-3.1$ are given in Tab.~\ref{Table:re4}.
\begin{table*}[htbp]
\centering
\renewcommand\arraystretch{1.2}
\caption{Pole positions $z_0$ and coupling constants $g_a$ for the $I=0$ states generated
in the $VB$ sector with $\mu=1$~GeV and $a(\mu)=-3.1$. \label{Table:re4}}
\begin{tabular}{p{2.5cm}<{\centering}p{2.5cm}<{\centering}p{2.5cm}<{\centering}p{2.5cm}<{\centering}}
\hline
\hline
\multirow{2}{*}{$z_0$ [MeV]}	&	\multicolumn{3}{c}{$g_a$} \\
&	$B^*\Xi_c$ & $B^*\Xi_c^\prime$ & $B_s^*\Lambda_c$	\\
\hline
$7768.81-2.5i$	&	$2.763+0.074i$	& 0	& $0.278-0.376i$ \\
&	2.764	&	0	&	0.468	\\
7894.43	& 0	& 2.037	&	0	\\
\hline
\hline
\end{tabular}
\end{table*}
Two poles are found in the complex energy plane, and they are just below the thresholds of
$B^*\Xi_c$ and $B^*\Xi_c^\prime$.
According to their positions together with the strengths of the couplings, these two poles
can be regarded as the $B^*\Xi_c$ and $B^*\Xi_c^\prime$ bound state, respectively.
For the pole at $z_0=7768.81-2.5i$~MeV, since the only considered decay pattern is $
B_s^*\Lambda_c$ here, the imaginary part of the pole position refers to half of the decay
width to this channel.
Thus, the decay width of this $\Lambda_{c\bar{b}}(7769)$ to $B_s^*\Lambda_c$ channel is about 5~MeV, and
the binding energy of this state is $26.7$~MeV.
For the $B^*\Xi_c^\prime$ bound state, there is no decay channel and it will naturally be observed
on the real axis, with a binding energy of about $9$~MeV.
Note  that the $PB$ and $VB$ interactions are calculated separately in this work.
The channels in these two sectors actually can interact with each other, and there will be
some changes in the results once all the channels are contained in one calculation.
For example, the $B^*\Xi_c^\prime$ bound state would obtain a decay width to $B\Xi_c^\prime$ channel,
and  the pole position would be slightly shifted.
However, this would not affect the existence of these dynamically generated states.
%

{\em Summary:}~~We have investigated the $B \Xi_c-B \Xi_c^\prime$ and $B^* \Xi_c-B^* \Xi_c^\prime$
systems via the exchange of a vector meson as the driving interaction in the
the Bethe-Salpeter equation (unitarized amplitude).
The possibility of dynamically generated poles has been examined for isospin $I=0$,
while for $I=1$ the potential is repulsive and can not form bound states.
In the $B \Xi_c-B \Xi_c^\prime$ interactions, four channels $B\Xi_c$, $B\Xi_c^\prime$, $B_s\Lambda_c$
and $B_c\Lambda$ are taken into consideration.
We find there are two poles located at $z_0=7720.12-3.7i$~MeV and $7847.60-2.2i$~MeV.
It can be seen that they are just below the threshold of $B\Xi_c$ and $B\Xi_c^\prime$, respectively,
and their coupling strength with each corresponding channel is much greater than that with other channels.
So these two poles can be regarded as $B\Xi_c$ and $B\Xi_c^\prime$  bound states.
The binding energies of these two states are about 30~MeV and 10~MeV.
We remark  that the quark content of these $\Lambda_{c\bar{b}}$ states is $[q\bar{b}][q^\prime sc]$,
where all the five involved quarks are different, a truely exotic hadron.
And due to this unique feature, its decay to light-meson–light-baryon channels are only possible
through the weak interaction, so that it is hard to observe them in the light-meson–light-baryon channels.
Thus, the dominant two-body decay channels of these states are $B_s\Lambda_c$ and $B_c\Lambda$,
and they are included in the coupled-channel calculation.
We find that  the decay width of the $B\Xi_c$ bound state is around $7.4$~MeV, and the $B\Xi_c^\prime$
bound state has a decay width of $4.4$~MeV.
For the $B^* \Xi_c-B^* \Xi_c^\prime$ interactions, the $B^*\Xi_c$, $B^*\Xi_c^\prime$ and $B_s^*\Lambda_c$
channels are considerd.
Like in the $PB$ case, the $B^*\Xi_c$ and $B^*\Xi_c^\prime$ channels each have a bound state,
whose locations are at $z_0=7768.81-2.5i$~MeV and $7894.43$~MeV, respectively.
In fact, to the order we are working, the $B^*\Xi_c^\prime$ channel does not interact with other channels.
Therefore, this $B^*\Xi_c^\prime$ bound state appears on the real axis, and it can not obtain
a width in this framework. More refined calculation to overcome this limitation are underway.
Finally, note that the existence of such $B^{(*)} \Xi_c^{(\prime)}$ bound states, which have the
most diverse flavor composition, would extend the already rich  spectrum of hadronic molecules.
These states can be searched for in the $pp\to B_c\Lambda X$ or
$pp\to B_s^{(*)}\Lambda_c X$ processes by the LHCb Collaboration in the $B_c\Lambda$ or
$B_s^{(*)}\Lambda_c$ invariant mass spectrum.
%

{\em Acknowledgments:}~~
We thank  Jia-Jun Wu and Bing-Song Zou for fruitful discussions.
This work is supported in part by the Deutsche Forschungsgemeinschaft (DFG)
and the National Natural Science Foundation of China (NSFC) through the funds provided to the
Sino-German Collaborative Research Center ``Symmetries and the Emergence of Structure in QCD"
(NSFC Grant No. 12070131001, DFG Project-ID 196253076 -- TRR 110).
The work of UGM was supported in part by The Chinese Academy of Sciences (CAS)
President’s International Fellowship Initiative (PIFI) (grant no. 2018DM0034) and
by the VolkswagenStiftung (Grant No. 93562).

\bibliographystyle{plain}

\clearpage
\pagebreak
\vspace{5cm}
\makeatletter
\setcounter{equation}{0}
\setcounter{figure}{0}
\setcounter{table}{0}
\setcounter{page}{1}
\renewcommand{\theequation}{S\arabic{equation}}
\renewcommand{\thefigure}{F\arabic{figure}}
\renewcommand{\thetable}{T\arabic{table}}
\renewcommand{\bibnumfmt}[1]{[S#1]}
\renewcommand{\citenumfont}[1]{S#1}

\section*{SUPPLEMENTAL MATERIAL}

\subsection{Lagrangians and potentials}

The Lagrangians for the $BBV$, $PPV$ and $VVV$ interactions used to evaluate
the processes under consideration are~\cite{Oset:2010tof}:
\begin{align}
  {\cal L}_{BBV} &= g \left(\langle \bar{B} \gamma_\mu [V^\mu, B]\rangle
  + \langle \bar{B} \gamma_\mu B \rangle\langle V^\mu\rangle\right)~, \nonumber \\
  {\cal L}_{VVV} &= i g \langle V^\mu [V^\nu, \partial_\mu V_\nu] \rangle~, \nonumber \\
  {\cal L}_{PPV} &= -i g \langle V^\mu [P, \partial_\mu P] \rangle~,
\end{align}
where $g=M_V/2f$ is the coupling constant in the hidden gauge formalism with the pion
decay constant $f \simeq 92$~MeV. For a general discussion of effective chiral Lagrangians
with vector mesons, see e.g. Ref.~\cite{Meissner:1987ge}.

Since the energy range we consider is close to the threshold, the three-momenta of all
the involved particles are assumed small comparing to their masses, except for the cases
when a heavy meson is exchanged.
Thus, we just keep the $\mu=\nu=0$ component during the calculation and the transition potential
for the $PB$ and $VB$ cases can, respectively, be simplified as follows (for a critical
discussion of this approximation, see~\cite{Gulmez:2016scm}):
\begin{align}
V_{PB\to PB} =& \frac{C_{ab}}{4f^2}(E_{P_1}+E_{P_2}), \nonumber \\
V_{VB\to VB} =& \frac{C_{ab}}{4f^2}(E_{V_1}+E_{V_2}) {\vec \epsilon_1}\cdot{\vec \epsilon_2},
\label{Eq:V}
\end{align}
where $a$ and $b$ are the channel indices for the initial and final states, respectively,
${\vec \epsilon}$ is the polarization vector of the external vector meson and the values of
the coefficients $C_{ab}$ for different isospin choices are given later when discussing each
specific case.

The potentials for heavy meson exchanges are different. These are the $D^*$ and $D_s^*$ exchange
in $B\Xi_c^{(\prime)} \to B_c\Lambda$ and $B_s\Lambda_c \to B_c\Lambda$ in this work.
The three-momenta of the initial and final particles can still be neglected, as they are on-mass-shell,
while the three-momentum transfer in the propagator of the exchanged $D^*$ and $D_s^*$ meson is different.
These exchanged heavy mesons can no longer be regarded approximately as on-shell any more.
Thus, the corresponding transition potential $V$ becomes:
\begin{equation}
V_{PB\to PB}^\prime = \frac{-C_{ab}g^2}{p_{ex}^2-m_{ex}^2}(E_{P_2}+E_{B_2}),
\end{equation}
where $p_{ex}^2=m_{P_2}^2+m_{B_2}^2-2E_{P_2}E_{B_2}$.

\subsection{Meson-baryon Green's function}
Next, we turn to the Green's function $G$, whose loop is formed by a meson ($P$) and a baryon ($B$).
Here, we apply  dimensional regularization to derive the following expression:
\begin{widetext}
\begin{align}
G^I =& i2M_B \int \frac{d^4q}{(2\pi)^4} \frac1{(P-q)^2-M_B^2+i\varepsilon} \frac1{q^2-M_P^2+i\varepsilon}
\nonumber \\
=& \frac{2M_B}{16\pi^2} \Big\{ a(\mu) + \ln\frac{M_B^2}{\mu^2} + \frac{M_P^2-M_B^2+s}{2s}
\ln\frac{M_P^2}{M_B^2} + \frac{\bar{q}}{\sqrt s} \big[ \ln(s-M_B^2+M_P^2+2\bar{q}\sqrt s)
+ \ln(s+M_B^2-M_P^2+2\bar{q}\sqrt s) \nonumber \\
&- \ln(-s-M_B^2+M_P^2+2\bar{q}\sqrt s) - \ln(-s+M_B^2-M_P^2+2\bar{q}\sqrt s) \big] \Big\},
\label{Eq:G}
\end{align}
\end{widetext}
where $P$ is the total four-momentum, $q$ is the four-momentum of the exchanged meson
and
$$\bar{q}=\sqrt{\frac{(s-(M_B+M_P)^2)(s-(M_B-M_P)^2)}{4 s}}$$
with ${\rm Im}(\bar{q})\ge0$.
The index $I$ of $G$ refer to the first Riemann sheet.
When considering the energy above the threshold for a certain channel, $G$ would change
into the second Riemann sheet for this corresponding channel.
Even though it appears that there are two parameters $a(\mu)$ and $\mu$ in Eq.~(\ref{Eq:G}),
they both appear only once and are summed together, so they can effectively be combined to one
free parameter.
In order to make a comparison with other works using the same method, here we always take the
regularization scale $\mu=1$~GeV and choose $a(\mu)$ to be a tunable parameter.
In Refs.~\cite{Wu:2010jys,Wu:2010vks}, the authors predict the existence of  hidden-charm pentaquark
states $N_{c\bar{c}}$ and $\Lambda_{c\bar{c}}$ with $a(\mu=1~\rm{GeV})=-2.3$.
This value is determined by comparing the results with another regularization of $G$ using a
sharp cutoff $\Lambda=0.8~\rm{GeV}$ in the three momentum.
Then they apply the same method to explore the hidden beauty system in Ref.~\cite{Wu:2010rvs}
using $a(\mu=1~\rm{GeV})=-3.71$ and several hidden beauty molecular states are predicted.
After the hidden-charm pentaquark states are observed by the LHCb
Collaboration~\cite{LHCb:2015yaxs,LHCb:2019keas}, Ref.~\cite{Xiao:2019aya} adopted this method
with $a(\mu=1~\rm{GeV})=-2.09$ to match with the experimental data.
Because the energy range studied in this work is between the energy of the hidden-charm and hidden
beauty sectors, the value of $a(\mu)$ should be taken between what are used in those two cases accordingly.
In the following discussions, we would take $a(\mu=1~\rm{GeV})=-3.1$ as a typical value.
Below, we will also give  results for $a(\mu)= 3.0$ and $a(\mu)=3.2$.

\subsection{Transition coefficients}

The coefficients $C_{ab}$ in the transition potential $V$ in Eq.~(\ref{Eq:V}) are
calculated following Ref.~\cite{Haacke:1975rt}. The $C_{ab}$ for $I=0$ and $1$ used in the
$PB$ sector are listed in Tab.~\ref{Table:IF0} and Tab.~\ref{Table:IF1}, respectively.

\begin{table}[hb]
\centering
\renewcommand\arraystretch{1.2}
\caption{Coefficients $C_{ab}$ used in the transition potential with $I=0$ in the $PB$ sector.
  \label{Table:IF0}}
\begin{tabular}{p{1.5cm}<{\centering}p{1.5cm}<{\centering}p{1.5cm}<{\centering}p{1.5cm}<{\centering}p{1.5cm}<{\centering}}
\hline
\hline
 & $B\Xi_c$ & $B\Xi_c^\prime$	& $B_s\Lambda_c$ & $B_c\Lambda$ \\
\hline
$B\Xi_c$ 	      & $-1$  & 0   & $-\sqrt2$  & $1/\sqrt2$ \\
$B\Xi_c^\prime$ &  	  & $-1$ 	& 0  & $-\sqrt{3/2}$ \\
$B_s\Lambda_c$ 	&     &     & 0  & 1 \\
$B_c\Lambda$   	&     &     &    & 0 \\
\hline
\hline
\end{tabular}
\end{table}

\begin{table}[hb]
\centering
\renewcommand\arraystretch{1.2}
\caption{Coefficients $C_{ab}$ used in the transition potential with $I=1$ in the $PB$ sector.
  \label{Table:IF1}}
\begin{tabular}{p{1.5cm}<{\centering}p{1.5cm}<{\centering}p{1.5cm}<{\centering}}
\hline
\hline
 & $B\Xi_c$ & $B\Xi_c^\prime$ \\
\hline
$B\Xi_c$ 	      & 1  & 0 \\
$B\Xi_c^\prime$ &  	 & 1 \\
\hline
\hline
\end{tabular}
\end{table}

The coefficients $C_{ab}$ for $I=0$ used in the $VB$ sector are
listed in Tab.~\ref{Table:IF02}.

\begin{table}[hb]
\centering
\renewcommand\arraystretch{1.2}
\caption{Coefficients $C_{ab}$ used in the transition potential with $I=0$ in the $VB$ sector. \label{Table:IF02}}
\begin{tabular}{p{1.5cm}<{\centering}p{1.5cm}<{\centering}p{1.5cm}<{\centering}p{1.5cm}<{\centering}}
\hline
\hline
 & $B^*\Xi_c$ & $B^*\Xi_c^\prime$	& $B_s^*\Lambda_c$ \\
\hline
$B^*\Xi_c$ 	      & $-1$  & 0   & $-\sqrt2$ \\
$B^*\Xi_c^\prime$ &  	  & $-1$    & 0 \\
$B_s^*\Lambda_c$ 	&     &     & 0 \\
\hline
\hline
\end{tabular}
\end{table}

\begin{table}[h!]
\centering
\renewcommand\arraystretch{1.2}
\caption{Pole position from each channel in $PB$ and $VB$ interactions for different values of $a(\mu)$.
  \label{Table:amu}}
\begin{tabular}{p{1.5cm}<{\centering}p{2.1cm}<{\centering}p{2.1cm}<{\centering}p{2.1cm}<{\centering}}
\hline
\hline
	& \multicolumn{3}{c}{$z_0$ [MeV]}	\\
	& $a(\mu)=-3.0$ &	$a(\mu)=-3.1$	&	$a(\mu)=-3.2$ \\
\hline
$B\Xi_c$ 	      	& 7748.82 & 7740.07 & 7718.42  \\
$B\Xi_c^\prime$ 	& 7857.32 & 7849.39 & 7828.83  \\
$B^*\Xi_c$  			& 7794.84 & 7785.25 & 7767.00  \\
$B^*\Xi_c^\prime$ & 7903.06	& 7894.43	& 7878.15  \\
\hline
\hline
\end{tabular}
\end{table}

\subsection{Further results}

To investigate the sensitivity of our results to the choice of subtraction constant,
we collect in Table~\ref{Table:amu} the predictions of the pole positions from each channel
for different values of the subtraction constant $a(\mu)$,
namely $-3.0$ and $-3.2$ (always setting $\mu=1\,$GeV).


\end{document}